\documentclass[a4paper,usenatbib]{mnras}

\usepackage{newtxtext,newtxmath}
\usepackage{amsmath}

\usepackage[T1]{fontenc}
\usepackage{ae,aecompl}

\usepackage{graphicx}	
\usepackage{amsmath}	
\usepackage{amssymb}	
\usepackage{soul}
\usepackage{multicol}

\title[Evidence against a supervoid causing the Cold Spot]{Evidence against a supervoid causing the CMB Cold Spot}

\author[R. Mackenzie et al.]{Ruari Mackenzie$^{1}$\thanks{E-mail: ruari.mackenzie@durham.ac.uk}, 
Tom Shanks$^{1}$, 
Malcolm N. Bremer$^{2}$, 
Yan-Chuan Cai$^{3}$,  \newauthor
Madusha L.P. Gunawardhana$^{1,4}$, 
Andr\'as Kov\'acs$^{5}$, 
Peder Norberg$^{1,6}$,
Istvan Szapudi$^{7}$\\
$^{1}$Centre for Extragalactic Astronomy, Department of Physics, University of Durham, South Road, Durham, DH1 3LE, UK\\
$^{2}$H.H. Wills Physics Laboratory, University of Bristol, Tyndall Avenue, Bristol BS8 1TL, UK\\
$^{3}$Institute for Astronomy, University of Edinburgh, Royal Observatory, Blackford Hill, Edinburgh EH9 3HJ, UK\\
$^{4}$Instituto de Astrof\'isica and Centro de Astroingenier\'ia, Facultad de F\'isica, Pontificia Universidad Cat\'olica de Chile, \\Vicu\~na Mackenna 4860, 7820436 Macul, Santiago, Chile\\
$^{5}$Institut de F\'isica d'Altes Energies, Barcelona Institute of Science and Technology, E-08193 Bellaterra (Barcelona), Spain\\
$^{6}$Institute for Computational Cosmology, Department of Physics, Durham University, South Road, Durham DH1 3LE, UK \\
$^{7}$Institute for Astronomy, University of Hawaii 2680 Woodlawn Drive, Honolulu, HI 96822, USA.}

\date{Accepted XXX. Received YYY; in original form ZZZ}

\pubyear{2017}

\begin{document}
\label{firstpage}
\pagerange{\pageref{firstpage}--\pageref{lastpage}}
\maketitle

\begin{abstract}
We report the results of the 2dF-VST ATLAS Cold Spot galaxy redshift
survey (2CSz)  based on imaging from VST ATLAS and spectroscopy from 2dF
AAOmega over the core of the CMB Cold Spot. We sparsely surveyed the
inner 5$^{\circ}$ radius of the Cold Spot to a limit of $i_{AB} \le
19.2$, sampling $\sim7000$ galaxies at $z<0.4$. We have found voids at
$z=$ 0.14, 0.26 and 0.30 but they are interspersed with small over-densities and
the scale of these voids is insufficient to explain the Cold Spot
through the $\Lambda$CDM ISW effect. Combining with previous data out to
$z\sim1$, we conclude that the CMB Cold Spot could not have been
imprinted by a void confined to the inner core of the Cold Spot.  Additionally we find that our `control' field GAMA G23 shows a similarity in its galaxy redshift distribution to the Cold Spot.
Since the GAMA G23 line-of-sight shows no
evidence of a  CMB temperature decrement we conclude that the Cold Spot
may have a primordial origin rather than being due to line-of-sight
effects.
\end{abstract}




\section{Introduction}
\label{sec:intro}

The Cosmic Microwave Background (CMB) provides the earliest snapshot of
the evolution of the Universe. Detailed observations of its structures
by the Wilkinson Microwave Anisotropy Probe (WMAP) and \emph{Planck}
missions have shown a universe broadly in concordance with the
$\Lambda$CDM paradigm. There remain a few anomalies which have been a
source of tension with standard cosmology and one such example is the CMB
Cold Spot (\citealp{Vielva}). The CMB Cold Spot is an $\sim5^{\circ}$
radius, -150 $\mu$K feature in the CMB in the Southern Hemisphere
which represents a departure arising in between	$<0.2$\%
(\citealp{Cruz}) to $<1-2$\% (\citealp{Planck2016CS}) Gaussian
simulations. It consists of a cold 5$^{\circ}$ radius core surrounded by
a less extreme 10$^{\circ}$ radius halo. The Cold Spot is also
surrounded by a high temperature ring which is important for its
original detection using a  Spherical Mexican Hat Wavelet (SMHW).

\begin{figure}
	\centering
	\includegraphics[width=\linewidth]{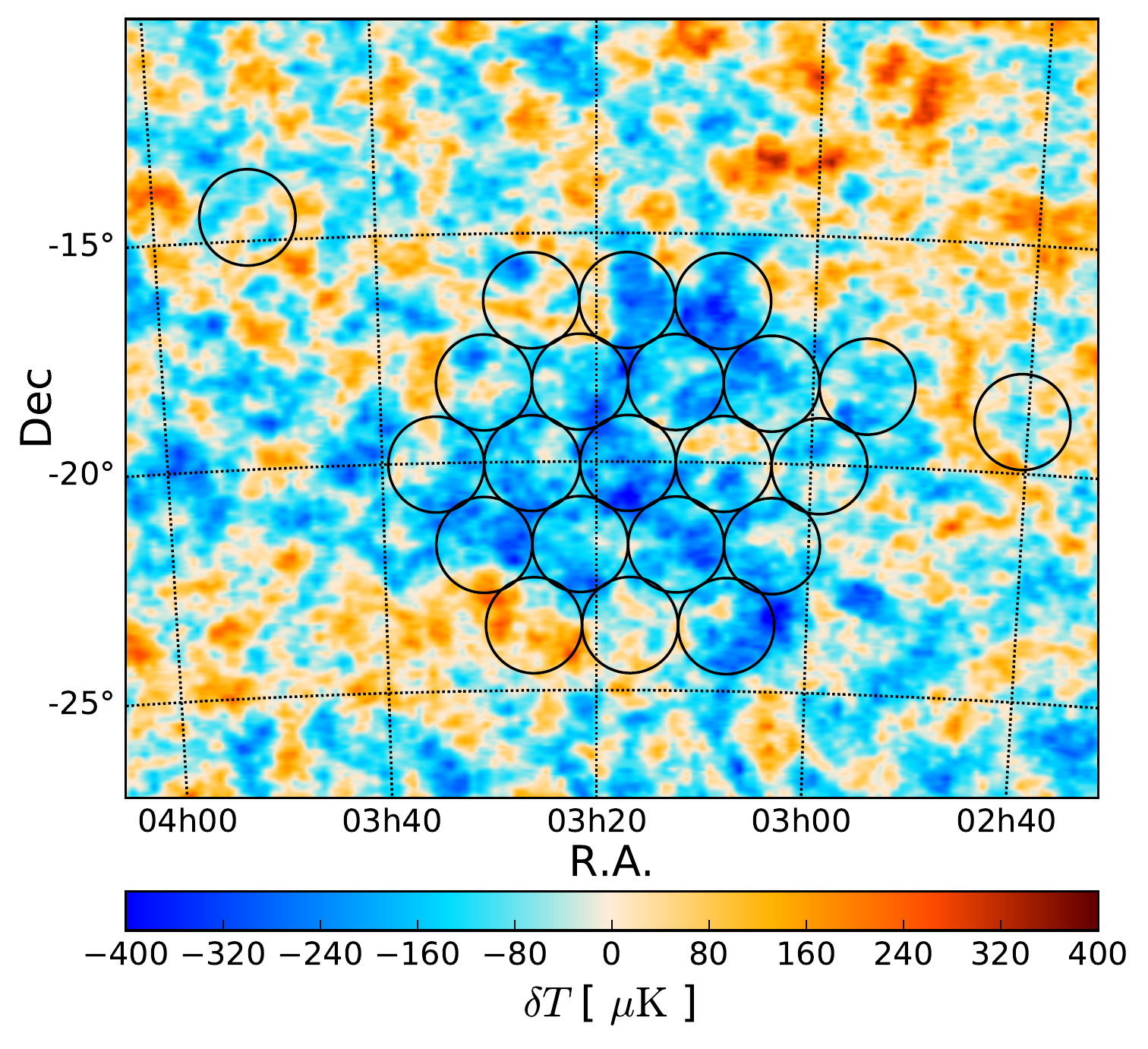}
	\caption{The 2CSz survey geometry: Superimposed on the Planck SMICA map of the CMB Cold Spot are circles representing  the 22 3 deg$^2$ galaxy redshift fields observed using AAT 2dF+AAOmega.  20 of these fields lie within a $5^{\circ}$ radius of the Cold Spot centre.}
	\label{fig:2df_map}
	\vspace{-1em}
\end{figure}

A number of proposals have been put forward to explain the Cold Spot,
including a non-Gaussian feature (\citealp{Vielva}), an artefact of
inflation (\citealp{Cruz}), the axis of rotation of the universe
(\citealp{Jaffe}) and the imprint of a supervoid via the Integrated
Sachs-Wolfe (ISW) effect (\citealp{InoueAndSilk}). The ISW effect \citep{SachsWolfe} occurs
in accelerating cosmologies due to the decay of gravitational potentials
over time. There is tentative statistical evidence to support the
existence of the ISW effect from the cross-correlation of large-scale
structure with the CMB, typically up to $3\sigma$ with single tracers and $4-4.5\sigma$ in some combined analyses (e.g. \citealp{CabreISW}, \citealp{Ho2008},
\citealp{Giannantonio}, \citealp{UtaneISW},
\citealp{GiannantonioStrikesBack}, \citealp{Planck2016ISW}). The ISW
effect must be measured statistically as the primary anisotropy
dominates on most scales. It has been hypothesized that a very large
void at $z<1$ could imprint itself on the CMB and explain the Cold Spot
in part (e.g. \citealp{InoueAndSilk}), however the ability of this to
explain the Cold Spot has been disputed (e.g. \citealp{Nadathur}). The
argument, prior to a detection of such a void, was that the probability
of any void occurring in $\Lambda$CDM was much lower than the
probability of the Cold Spot arising from primordial Gaussian
fluctuations.

The significance of the Cold Spot as an anomaly has been widely
discussed. The main problem is to quantify the amount of   {\it a
posteriori} selection in the  originally claimed 0.2\% significance of 
\citet{Cruz}. In particular, \citet{ZhangHuterer} pointed out that
the use of top-hat or Gaussian kernels provided much lower
significance for the Cold  Spot than the original SMHW kernel and
\citet{Bennett} emphasised this viewpoint in their review.
\citet{Vielva2010} argued that as long as the original Cold Spot
detection was `blind' and the SMHW kernel well-motivated in a search for
non-Gaussian features then this `look elsewhere' effect in terms of
kernels was less relevant. \citet{Zhao}, \citet{Gurzadyan} and
\citet{Planck2016CS} tried a related approach to address the Cold
Spot significance and chose the coldest pixels in CMB simulations to
look at the small-scale statistics within the surrounding pixels. In the
version of \citet{Planck2016CS}, it was found that the temperature
profile of the Cold Spot was poorly described by the simulations with
$<1-2$\% having a higher $\chi^2$ compared to the mean than the data.
Here we shall essentially adopt this approach, now following
\citet{Nadathur} and \citet{Naidoo}, and ultimately test how much
any foreground void that is found can reduce this 1-2\% significance
assuming the original SMHW kernel.

Motivated by theoretical discussion there have been many 
attempts to detect a void associated with the CMB Cold Spot.
\cite{Rudnick} searched NVSS radio sources and claimed to find a lower density of objects in the Cold Spot region but this was disputed by \cite{SmithHuterer}. 
\cite{Granett2010} used 7 CFHT MegaCam fields to make a photo-$z$ 
survey for large under-densities. They found no evidence of a void $0.5<z<0.9$ but their data was consistent with a low-$z$ void.
This was in line with \cite{FrancisPeacock} who found evidence for an under-density in 2MASS in the Cold Spot direction but the ISW imprint was $\sim5\%$ of the CMB Cold Spot temperature decrement. \cite{Bremer} used VLT VIMOS to make a $21.9< i_{AB}<23.2$ galaxy redshift survey in 6 small sub-fields of the Cold Spot area. The total area covered was $0.37$deg$^2$ and the redshift range  covered was $0.35 < z < 1$. Using VVDS \citep{VVDS} data as control fields, \cite{Bremer} found no evidence for anomalously large voids in the Cold Spot sightline.
At lower redshifts, \cite{Szapudi} using a Pan-STARRS, 2MASS and WISE combined catalogue,
constructed photometric redshifts and detected a 220h$^{-1}$Mpc radius supervoid
with a central density contrast, $\delta_{m}\sim -0.14$,  spanning $z \approx 0.15 -
0.25$. However, this supervoid would not explain the entirety of the CMB
Cold Spot as a $\Lambda$CDM ISW effect. The authors argued that the
alignment of the Cold Spot and the supervoid could be evidence of a
causal link due to some mechanism beyond standard cosmology. It has been argued that there is evidence for voids showing an ISW-like effect above the standard prediction (e.g. \citealp{Granett}) but at marginal significance and other analyses have found results consistent with standard cosmology (e.g. \citealp{NadathurISW}, \citealp{Hotchkiss}).
\cite{Kovacs} extended this work to include photometric redshifts from 2MASS (2MPZ)
and spectroscopic redshifts from 6dFGS. Using these datasets it was
claimed that the under-density detected by \cite{Szapudi} extends along
the line of sight back to $z\sim0$ with a void radius of up to $500 $
h$^{-1}$Mpc. The void was suggested to be elongated in the redshift
direction and had a smaller radius of $195$h$^{-1}$Mpc in the
angular direction. Even  with these larger estimates of the $z\approx 0.15$ 
void's scale the Cold Spot temperature may only be partly  explained by the
$\Lambda$CDM ISW effect. But significant uncertainties remain in the void
parameters due to the nature of photometric redshifts, and in order to test
claims of divergence from $\Lambda$CDM, the parameters of the supervoid
must be better determined. The sightline must also be unique in order
to explain the uniqueness of the Cold Spot in the CMB. 

We have therefore carried out the 2dF-VST ATLAS Cold Spot Redshift Survey (2CSz) over the inner
5$^{\circ}$ radius core of the Cold Spot in order to test the detection
made by \cite{Szapudi} and, if the supervoid were confirmed, to measure
its parameters to assess any tension with $\Lambda$CDM. Throughout the
paper we use Planck 2015 cosmological parameters
(\citealp{Planck2016cosmo}), with $H_0=100$h kms$^{-1}$Mpc$^{-1}$, $h$=0.677,
$\Omega_{M,0}$=0.307, $T_{cmb,0}$=2.725 K.

\section{Survey and Data Reduction}

\begin{figure*}
	\centering
	\includegraphics[width=\linewidth]{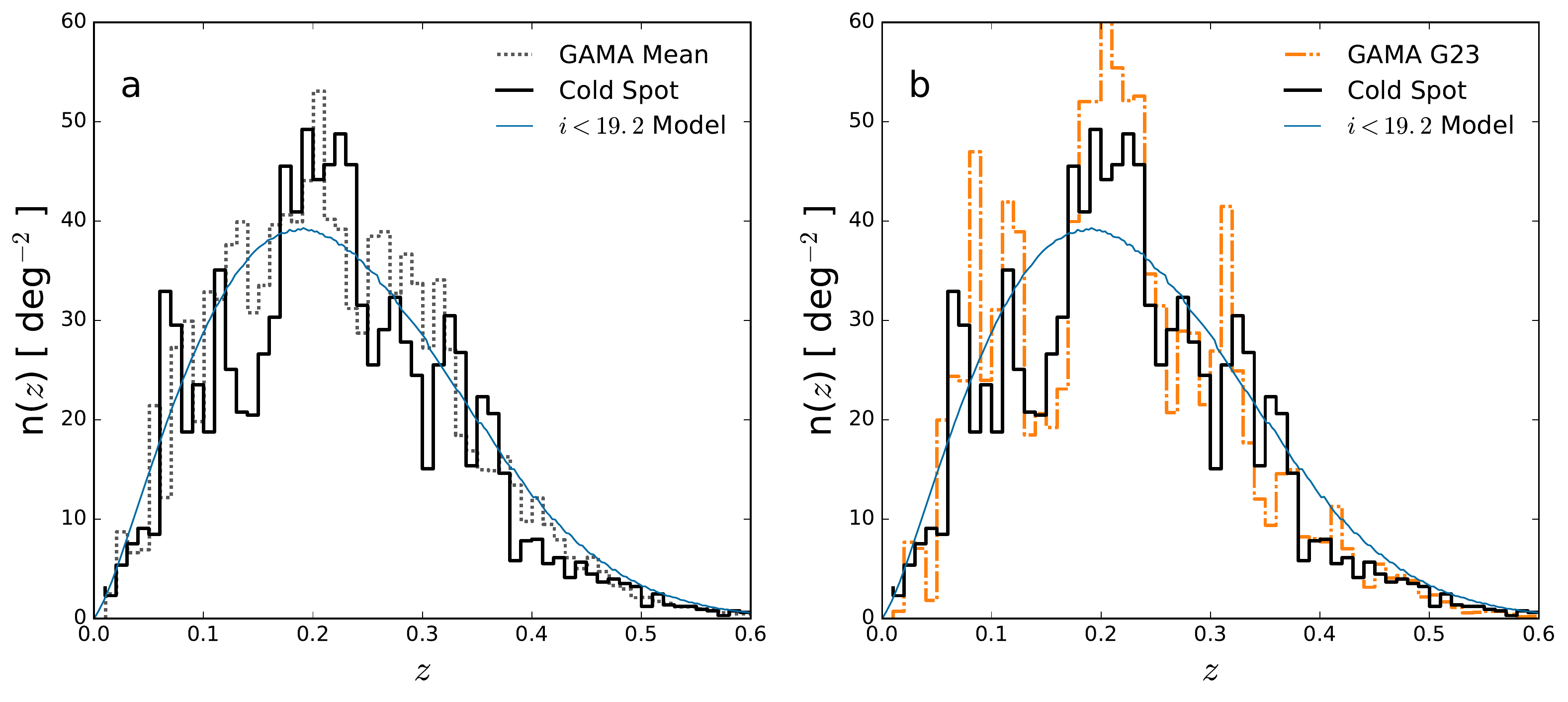}
	\caption{(a) The galaxy redshift distribution of the 2CSz (black).
	Also shown is the $n(z)$ from the average of the 4 GAMA fields at 
	R.A. $\sim$ 9h, 12h, 15h and 23h (G23) at the same $i_{AB}<19.2$ limit (grey dotted) and the
	 homogeneous model of \citep{Metcalfe} (blue). (b) The galaxy redshift distribution of the
	  2CSz (black). Also shown is the $n(z)$ from the GAMA G23 field, at the same $i_{AB}<19.2$ limit (yellow dot-dashed), and the same homogeneous model as in (a) (blue).}
	\label{fig:redshift_dist}
	\vspace{-1em}
\end{figure*}

The first goal of 2CSz was to probe the supervoid of \cite{Szapudi} with spectroscopic precision. We therefore targeted the inner 5$^{\circ}$ radius with 20 contiguous 2dF fields (see Fig. 1). A further 2 fields were targeted at larger radii in the sightlines of two $z\sim0.5$ quasars, which, in other work, will be used with HST COS spectra to probe the void structure in the Lyman $\alpha$  forest as well as in the galaxy distribution. In all fields, 2dF  galaxies were sampled at a rate of $\sim$110 deg$^{-2}$ to a limit of $i_{AB}<19.2$. The survey was selected analogously to the GAMA G23 survey\footnote{
	The Galaxy And Mass Assembly (GAMA) survey
	(\citealp{GAMA}, \citealp{Driver11}) includes 3 Equatorial fields at R.A. $\sim$ 9hrs, 12hrs and 15hrs,
	each covering about 60 deg$^2$, highly spectroscopically complete to $r_{AB} < 19.8$. There is also one
	SGC field (G23) covering 50 deg$^2$ similarly complete to $i_{AB} = 19.2$ \citep{GAMALiske}.
	}, but sub-sampled to the number density matched to
a single 2dF pointing per field ($\sim1/8$ sampling). This provided us
with a highly complete control field.

The imaging basis for this spectroscopic survey was the VLT Survey
Telescope (VST) ATLAS (\citealp{ATLAS}), an ongoing $\sim$4,700deg$^2$
$ugriz$ survey at high galactic latitude over the two sub-areas in the North and South Galactic Caps (NGC and SGC respectively), the latter of which includes the Cold Spot region. VST ATLAS
reaches an $i$ band 5$\sigma$ depth of 22.0 AB mag for point sources and
has a median $i$ band seeing of $0.81''$, allowing clean star-galaxy
classification to our magnitude limit. The main selection criterion was
to select extended sources with $i_{Kron, AB} \le 19.2$ where $Kron$
indicates a pseudo-total magnitude with the usual definition. Additional
quality control cuts were applied to the data to ensure the removal of
stars and spurious objects from the galaxy catalogue. Although the
extended source classification removes most stars, an additional
star-galaxy cut was applied ($i_{Kron, AB} - i_{ap3, AB} < 0.1\times i_{Kron, AB}-1.87$)
where $i_{ap3, AB}$ denotes the magnitude corresponding to the flux within a
$2''$ diameter aperture (c.f. Fig. 22 of \citealp{ATLAS}). To reject
spurious objects (e.g. ghosts around bright stars), sources without $z$
band detections were rejected, as were objects near Tycho-2 stars at
radii calibrated to VST ghosts. Additionally, a cut of SKYRMS $ \le 0.2$ ADU
was applied to the RMS of the sky measurement for each source in the
catalogue to  remove further artefacts. These cuts were validated with
GAMA G23. All magnitudes were corrected for Galactic extinction
(\citealp{PlanckDust}).

The spectroscopic survey was completed in 22 2dF fields with 20 covering 
the inner 5$^{\circ}$ radius of the Cold Spot. The survey footprint is
shown in Fig.  \ref{fig:2df_map}. 2dF covers a 3 deg$^2$ area with
approximately 392 fibres, $\sim25$ of which were used as sky fibres. The
number density of selected galaxies was 722 deg$^{-2}$, further
randomly sampled down to $\sim200$deg$^{-2}$ in order to provide
sufficient targets to utilise all fibres. Many targets cannot be
observed due to limitations in positioning of the fibres to avoid fibre
collisions and to limit fibre crossings. This down-sampled target list
was finally supplied to the 2dF fibre allocation system \emph{Configure}. 

The spectroscopic observations were carried out in visitor mode on
$16^{th}$, $17^{th}$ and $18^{th}$ of November 2015, during grey (Moon phase)
conditions with typical seeing of $\sim2.0''$. We observed using AAOmega with
the 580V and 385R gratings and the 5700\AA~ dichroic. This gives a
resolution of $R\sim1300$ between 3700\r{A}  and 8800\AA. Each field was
observed with $3\times15$ minute exposures; flats and an arc frame were also
taken with each plate configuration. Fields observed at
high airmass at the beginning and end of the night had additional 15
minute exposures where possible. Dark and bias frames were taken during
the day before and after each night. 

Spectroscopic observations were reduced and combined using the $2dFdr$
pipeline (\citealp{2dfdr}, \citealp{2dfdr2010}). The data was corrected with the fibre flat and median sky
subtracted. Dark frames were not ultimately used as on
inspection they did not improve the data quality. The sky correction
parameters used were throughput calibration using sky lines, iterative
sky subtraction, telluric absorption correction and PCA after normal sky
subtraction. The resulting reduced spectra were then redshifted manually
using the package $runz$ \citep{runz}. Redshifts were ranked in quality
from 5 (Template quality), 4 (Excellent), 3 (Good), 2 (Possible) and 1
(Unknown redshift). Only redshifts of quality 3 or greater were used in
the final science catalogue. Typically excellent quality redshifts had
multiple strong spectra features (e.g. H$\alpha$, [OII] and Ca II K and
H lines) and good redshifts contained at least one unambiguous feature.
Overall the redshift success rate was approximately $89\%$ ranging from
$71\%$ to $97\%$; typically the success rate is a strong function of the
phase and position of the Moon. 

With an  $89\%$ success rate, incompleteness will have only a small
effect if redshift failures are random rather than systematic and we
modeled this with GAMA G23. To test what effect magnitude dependent
completeness could have on these results we measured the completeness
with magnitude for our survey, finding  that completeness is $\sim96\%$
for $i_{AB}\le18.2$ and decreases to $\sim82\%$ for $18.7<i_{AB}\le19.2$. This
magnitude dependent completeness will bias the $n(z)$ towards the
redshift distribution of the brighter galaxies. To estimate the effect
this has on the $n(z)$ we weight the GAMA G23 $n(z)$ with the
completeness as a function of magnitude from 2CSz. Taking the ratio of
the weighted and unweighted $n(z)$ we obtain the completeness fraction
as a function of redshift,  $f(z)\approx 0.95-0.232z$ for $z<0.45$. This
linear modulation of the $n(z)$ does not significantly affect the
results but this analysis assumes that redshift failures depend only on
the magnitude of the object and not the redshift. We do not apply a 
correction to the data as we do not believe this assumption holds 
(see Section \ref{systematics}).

\begin{table}
	\begin{center}
		\begin{tabular}{cc}
			\hline
			Main Selection & $i\le19.2$  \\
			Area & $66$ deg$^2$  \\
			Number of Galaxies* & $6879$  \\
			Completeness & $89\%$ \\
			Redshift Range & $0.0\le z\le0.5$  \\
			Galactic coordinates ($l,b$) & (209, -57) \\
			\hline
		\end{tabular}
		\begin{tabular}{cl}
			* & 	Galaxies with redshift quality $\ge$ 3
		\end{tabular}
		\caption{2CSz survey parameters. \label{tab:survey_params} }
	\end{center}
\end{table}

\section{Results}
\subsection{Redshift Distributions}

The 2CSz redshift distribution of the $\sim$6879 quality $>2$ galaxies is
shown in Fig. \ref{fig:redshift_dist}(a), along with the mean GAMA redshift
distribution and a homogeneous model \citep{Metcalfe}. Comparison with the homogeneous model allows for under and over-densities to be identified. 
Due to the sub-sampling of the spectroscopic survey we normalised the
$n(z)$'s to the galaxy number magnitude counts in the Cold Spot and G23
regions using an ATLAS $iz$ band-merged catalogue. 
We found that the 75deg$^2$ Cold Spot area was $16\pm3$\% under-dense relative to the $\sim1000$deg$^2$ around G23. We also found that the Cold Spot had a $7.4\pm0.7$\% number density deficit relative to a similarly large $\sim1000$deg$^2$ region surrounding the Cold Spot whereas the G23 galaxy count was consistent with the SGC average over its full $\sim2600$deg$^2$ area. Both the SGC number count and the mean galaxy density averaged over the  4 complete GAMA fields are in good agreement with the homogeneous model. To allow comparison with G23 we chose to normalise the Cold Spot observed $n(z)$ by 7.4\% lower in total counts than both the homogeneous model and the G23 observed $n(z)$ and this is what is shown in Fig. 2. Ignoring the large scale gradient like this is certainly correct if it is a data artefact. But there is also a case to be made for it even if it is real since the Cold Spot is essentially a small-scale, $\sim75$deg$^2$, feature rather than a $\sim1000$deg$^2$ feature.

Here and throughout field-field errors are used. These are based on a (2dF) field size of $\sim 3$ deg$^2$.

The mean GAMA redshift distribution comes from
the 4 GAMA fields, G23, G09, G12 and G15 selected with $i_{AB} \le 19.2$. The
latter three $r$-limited fields were checked to be reasonably complete at
the $i_{AB} \le 19.2$ limit for this analysis. The
stacked GAMA redshift distribution fits well with the \cite{Metcalfe}
homogeneous model for galaxies with $i_{AB}\le19.2$. Fig.
\ref{fig:redshift_dist}(a) shows indications of inhomogeneity in the Cold Spot
sightline where we see evidence of an under-density spanning $0.08<z<0.17$
and there is also a hint of a smaller under-density at $0.25<z<0.33$. This
would be consistent with the \cite{Szapudi} supervoid but we also see
evidence for an over-density at $0.17<z<0.25$, apparently in conflict with the
previous claim that the supervoid was centred in this range. Given the
photometric redshift error, there may be no real contradiction between
the datasets but their single void model does appear inconsistent with
our spectroscopic data (see Section \ref{photoz}). Another under-density is seen at $0.37<z<0.47$ but
systematic errors, such as spectroscopic incompleteness, become more 
important at this point (see Section \ref{systematics}).

\subsection{Void Model}

In order to obtain the parameters of an under-density and determine its
ISW imprint a void profile must be selected and fit to the redshift
distribution. Some previous work has used simple top-hat void models
as the measured profile was dominated by photo-$z$ error. In the case of
our well sampled spectroscopic survey the structure of the void is
important to the fitting and allows us to estimate the ISW imprint of
any void. Following \cite{Kovacs}, we have chosen the $\Lambda$LTB void profile
described by a Gaussian potential (i.e. $\alpha = 0$ in \citealp{Finelli}, eq.
\ref{eqn:void_density}) which will allow us to use the analytic
expression for the ISW temperature profile given by these authors. 
This compensated void profile is described by eq. \ref{eqn:void_density},

\begin{equation}
\centering
\delta_m(r) = \delta_0\ g(a) \left( 1 - \frac{2}{3}\,\frac{r^2}{r_0^2} \right) \exp\left(-\frac{r^2}{r_0^2}\right)  ,
\label{eqn:void_density}
\end{equation}

\noindent where $\delta_m(r)$ is the matter density contrast at radius $r$
from the void centre, $\delta_0$ is the matter density contrast at the void centre, $g(a)$ is the  growth factor at scale factor, $a$, and $r_0$ is the void
radius. As shown by \cite{Finelli} the ISW imprint of a void described by
eq. \ref{eqn:void_density} can be calculated using eq.
\ref{eqn:dT},

\begin{equation}
\centering
\begin{split}
 \frac{\delta T}{T}(\theta) \approx \frac{3 \sqrt{\pi}}{22} \frac{H(z_0) \Omega_{\Lambda} F_4(-\Omega_{\Lambda}/\Omega_{M}(1+z_0)^3)}{H_0 (1+z_0)^4 F_1(-\Omega_{\Lambda}/\Omega_{M})} \times \\
 \left( 1+\text{erf}\left(\frac{z_0}{H(z_0) r_0}\right) \right) \delta_0 (H_0 r_0)^3 \exp \left[-\frac{r^2(z_0)}{r_0^2}\theta^2\right],
 \end{split}
\label{eqn:dT}
\end{equation}

\noindent where $\frac{\delta T}{T}(\theta)$ is the ISW temperature
imprint at angle $\theta$ away from the centre of the void and $z_0$ is 
the central redshift of the void. $F_1$ and $F_4$ are described by eq.
\ref{eqn:F1} and \ref{eqn:F4} respectively where ${}_2 F_1$ is the
hypergeometric function,

\begin{equation}
F_1 = \ {}_2 F_1\ \left[1,\frac{1}{3},\frac{11}{6},\frac{-\Omega_{\Lambda}a^3}{\Omega_M}\right], \label{eqn:F1} 
\end{equation}
\begin{equation}
F_4 =  \ {}_2 F_1\ \left[2,\frac{4}{3},\frac{17}{6},\frac{-\Omega_{\Lambda}a^3}{\Omega_M}\right].
\label{eqn:F4}
\end{equation}

\cite{Finelli} also give an expression for the Rees-Sciama effect \citep{ReesSciama}, the
second order ISW effect. As the Rees-Sciama effect is sub-dominant to the ISW effect at the scale of the CS at low redshift in the standard cosmology  \citep{Cai2010} , we will neglect its contribution in our calculations. 

\begin{figure*}
	\centering
	\includegraphics[width=\linewidth]{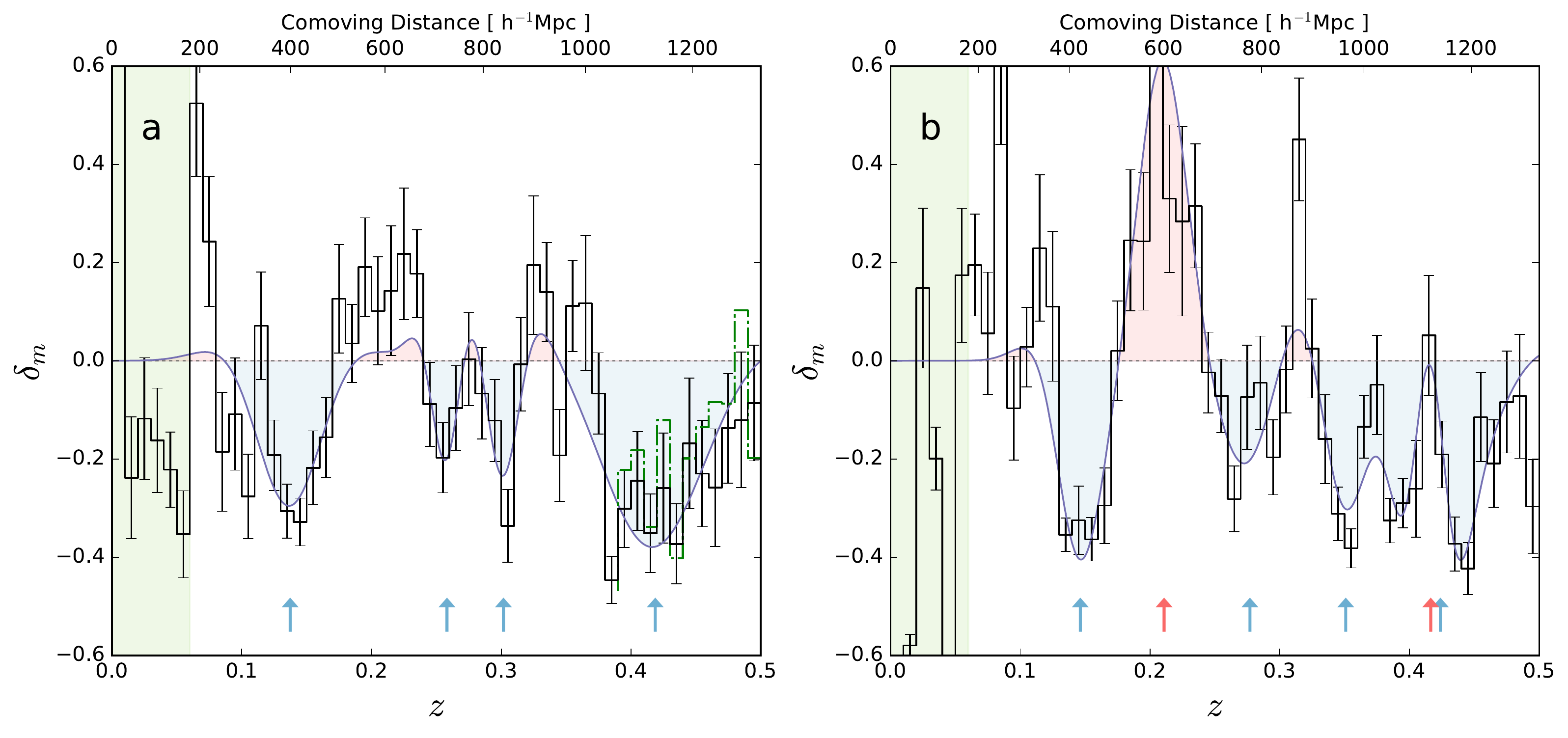}
	\caption{(a) The matter density contrast for 2CSz (black histogram), the
	best-fit  void models (dark blue) and the `Local Hole' extent (green),
	modelled under-densities are filled in blue and overdensities in
	red. The dashed line shows the result at $z>0.38$ when only 2dF
	fields with $>90$\% redshift success rate are used.  Arrows indicate
	the centre of each fitted under-density (blue) and over-density (red).
	(b) The mass density contrast for the GAMA G23 with symbols as in
	(a).}
	\label{fig:void_fit}
	\vspace{-1em}
\end{figure*}

\subsection{Perturbation Fitting in the Cold Spot}

In order to estimate the ISW imprint of the observed inhomogeneities we
have fitted the redshift distribution with compensated perturbations with
the profile given by eq. \ref{eqn:void_density}. Although our spectroscopic 
survey has 3D information we pursue this 1D analysis to mimic the void finding 
used in past photo-$z$ analyses, so the same large under-densities are selected. In order to do this it
was first necessary to transform the $n(z)$ to the matter density
contrast, $\delta_m(z)$, done by first converting to the galaxy density contrast, $\delta_g(z)$, and then dividing by the galaxy bias, $b_g$. 
These transformations are shown in eq. \ref{eqn:galaxy_density_contrast} and \ref{eqn:matter_density_contrast},

\vspace{-1cm}
\begin{multicols}{2}
	\begin{equation}
	\centering
	\delta_g(z) = \frac{n(z)}{n_{\mathrm{model}}(z)}-1 
	\label{eqn:galaxy_density_contrast}
	\end{equation}\break
	\begin{equation}
	\delta_m(z) = \frac{\delta_g(z)}{b_g}  
	\label{eqn:matter_density_contrast}
	\end{equation}
\end{multicols}

\noindent where $n_{model}(z)$ is the predicted redshift distribution from the homogeneous model \citep{Metcalfe}. Since the magnitude limits for the 2CSz and G23 galaxies are the same,  the bias for both samples can be estimated from the GAMA G23 correlation function, obtaining a linear bias of $b_g = 1.35$. Although simplistic, this linear bias assumption is accurate enough for the large scales of interest here.

Fig. \ref{fig:void_fit}(a) shows the matter density contrast for the 2CSz
survey, assuming field-field errors. A number of features can be seen in
Fig. \ref{fig:void_fit}(a). At the lowest redshifts ($z<0.06$) the `Local
Hole' can be seen as a $\sim25\%$ under-density. This is well studied in
the literature and seems to extend across the SGC (e.g
\citealp{WhitbournandShanks}). At $z=0.06$ there is an over-density
separating the `Local Hole' from a $\sim40\%$ under-density which
extends to $z=0.17$. Another peak in the distribution is followed by two
under-densities ($z=0.23$, 0.25 and 0.3 respectively). Lastly there is
a clear break at $z=0.38$ and a $\sim30\%$ under-density extending to
$z=0.5$ where it converges towards the homogeneous model. This feature
may be due to redshift dependent incompleteness as we will discuss
later (see Section \ref{systematics}).

In order to fit the redshift distributions in an unbiased way we have
adopted an iterative fitting procedure that minimises the necessary
complexity of any fit, quantified with the Akaike information criterion
(AIC) (e.g. \citealp{AIC}). The AIC statistic takes into account the improvement in the fit of a more complex model but additionally penalizes it for this increased complexity. We use the AIC statistic specifically because it can be corrected in the case  when the number of data points is not much larger than the number of parameters.
We have fitted individual under-densities, $\delta_m(r)$, with 3D perturbations described by eq.
\ref{eqn:void_density}, averaged over the 5 deg radius of 2CSz. In order
to describe the features seen in Fig. \ref{fig:void_fit} we model the
line of sight $n(z)$ as a combination of perturbations. The fitting
assumes the void is centred on the Cold Spot. The whole redshift range
was fitted simultaneously, with the `Local Hole', at $z \le 0.0625$
excluded from the fit as it is not unique to the Cold Spot. 
We do not believe this will affect our results as there is a clear 
over-density, which appears to be a wall, separating the Local Hole and the lowest redshift void we consider. 
Our iterative method initially assumed N perturbations seeded with random
parameters and fitted them to the data. Fitting was carried out with a Levenberg-Marquardt algorithm and quoted errors are standard errors calculated from the covariance matrix.
 Iterating over new random values and fitting we converge on the best fit parameters for N perturbations. 
The best fits for each value of N were then compared via the corrected AIC statistic, the
minimum of which gave the optimum fit and the relative likelihood
allowed for other values of N to be rejected if significantly poorer. The corrected AIC statistic is given by eq. \ref{eqn:AIC} (\citealp{AIC}) where $k$ is the number of parameters being fit, $N_{\mathrm{data}}$ is the number of data points and $\hat{L}$ is the maximised likelihood function, 

\begin{equation}
\begin{split}
\label{eqn:AIC} 
\mathrm{AIC} & = 2k - 2 ln(\hat{L}) + \frac{2k(k+1)}{N_{\mathrm{data}}-k-1} \\
 & = 2k + \chi^2 + \frac{2k(k+1)}{N_{\mathrm{data}}-k-1}.
\end{split}
\end{equation}

\noindent The second line of eq. \ref{eqn:AIC} holds in the case of normally distributed residuals. The relative probability of one model over another with a greater AIC value is given by the Akaike weights (eq. \ref{eqn:AIC_p}) where $\mathrm{AIC}_{\mathrm{min}}$ is the minimum AIC, $\Delta \mathrm{AIC}_i = \mathrm{AIC}_i-\mathrm{AIC}_{\mathrm{min}}$ and $k_{\mathrm{max}}$ is the maximum k considered,

\begin{equation}
\label{eqn:AIC_p} 
w_i = \frac{ e^{-\Delta \mathrm{AIC}_i/2}}{\sum_{k=1}^{k_{\mathrm{max}}}e^{-\Delta \mathrm{AIC}_k/2}}.
\end{equation}

\noindent Hence a $p=0.05$ rejection of the weaker model corresponds to a $\Delta \mathrm{AIC}\sim 6$  and we shall adopt $\Delta \mathrm{AIC}= 6$ as a threshold for rejecting models over the best fit. More complex models were considered until one was rejected over a simpler model. 

This analysis suffers from degeneracies in that we cannot discern
the difference between two voids and a wide void with an interior, narrow
over-density. For this reason, the fitting ranges of parameters were
restricted in range to provide sensible fits. Specifically we restricted the 
void radius to be between 50 and 150 h$^{-1}$Mpc and the central density 
contrast was constrained to lie in the physical range, $\delta_0\ge-1$. 
Parameters at the radius limits were individually re-fit. Fits were also rejected which had perturbations at the very edges of the fitting range, i.e. $z_0 < z_{min} +0.01$ or $z_0 > z_{max} -0.01$. Additionally the compensated
profile we have adopted cannot describe sharp narrow under-densities as
they are averaged out in the survey field; however the purpose of this
analysis is to detect large voids and this places upper limits on ISW
contributions. We have allowed over-densities to be fitted with the
perturbation described by eq. \ref{eqn:void_density} but as this
profile was derived for voids the resulting $\delta T$ values should be
treated with caution. The minimum AIC values for each value of N perturbations are shown in Table \ref{tab:AIC}. The resulting best fits are shown in Fig.
\ref{fig:void_fit}. For the Cold Spot, the iterative procedure selected
N$=4$ perturbations as the best fit (all under-densities) to give the fits summarized in
Table \ref{tab:void_params}. The AIC test does not strongly reject the N$=5$ solution but we note the difference between the models is only in the fitting of the $z\sim0.42$ void with one profile or two and the resulting total $\delta T$ differs by just $2.7 \mu K$ which is not significant. 

\begin{table}
	\begin{center}
		\begin{tabular}{cccc}
			\hline
			$N$ & $k$ & \multicolumn{2}{c}{$\mathrm{AIC}_{\mathrm{min}}$} \\ \hline
			    &     & Cold Spot & G23    \\ \hline
			1   & 3   &  248.85   & 441.97   \\
			2   & 6   &  147.17   & 240.14   \\
			3   & 9   &  131.65   & 197.66   \\
			4   & 12  &  (123.91)   & 154.64   \\
			5   & 15  &  125.18   & 151.07   \\
			6   & 18  & 132.33 & (141.45)   \\
			7   & 21  &   -       & 149.86   \\ 
			\hline 
		\end{tabular}
		\caption{The minimum AIC values for each value of N perturbations for the Cold Spot and G23. $k$ is the number of free parameters. The minimum AIC values best fits for are shown in parenthesis.}\label{tab:AIC}
	\end{center}
\end{table}

\begin{table}
	\begin{center}
		\begin{tabular}{ccccc}
			\hline
			 & $z_0$ & $r_0$  & $\delta_0$ & $\delta T (\theta=0)$  \\
			 &    &(h$^{-1}$Mpc)&           & ($\mu$K) \\
			\hline
			Cold Spot &&&& \\
			\hline
			& 0.14$\pm$0.007 & 119$\pm$35 & -0.34$\pm$0.08 & -6.25$\pm$5.7 \\
			& 0.26$\pm$0.004 & 50$\pm$13 & -0.87$\pm$0.12 & -1.02$\pm$0.8 \\
			& 0.30$\pm$0.004 & 59$\pm$17 & -1.00$+$0.72 & -1.80$\pm$2.1 \\
			& 0.42$\pm$0.008 & 168$\pm$33 & -0.62$\pm$0.16 & -22.6$\pm$14.7 \\
			\hline
			G23 &&&& \\
			\hline
			& 0.15$\pm$0.004 & 82$\pm$33 & -0.49$\pm$0.17 & -2.92$\pm$3.7 \\
			& 0.21$\pm$0.006 & 88$\pm$21 & +0.89$\pm$0.35 & +6.09$\pm$5.1 \\
			& 0.28$\pm$0.007 & 85$\pm$29 & -0.36$\pm$0.24 & -2.06$\pm$2.6 \\
			& 0.35$\pm$0.006 & 74$\pm$22 & -1.00$+$0.10 & -3.40$\pm$3.1 \\
			& 0.42$\pm$0.005 & 150$\pm$20 & -0.63$\pm$0.13 & -16.1$\pm$7.4 \\
			& 0.42$\pm$0.002 & 50$\pm$5 & +4.16$\pm$1.6 & +3.96$\pm$2.0 \\
			\hline
		\end{tabular}
		\caption{Best fit 3-D $\Lambda$LTB parameters for compensated perturbations (eq. \ref{eqn:void_density}) estimated from the Cold Spot and GAMA G23 density
contrast profiles in Fig. \ref {fig:void_fit}. The central temperature
decrement, $\delta T$, predicted from the ISW effect is also given. 
\label{tab:void_params}}
	\end{center}
\end{table}

\subsection{Perturbation Fitting in GAMA G23}

As noted above, we originally planned to use GAMA G23 as a control field
but analysis showed that even on 50 deg$^2$ scales there was sufficient sample variance to
merit using a model which we validated with the stacked $i_{AB}\le19.2$
$n(z)$ from all four GAMA fields with a combined area of $\sim$240deg$^2$. 
Indeed, Fig. \ref{fig:redshift_dist}(b) shows
that upon comparison the Cold Spot redshift distribution bears
remarkable similarity with G23 in the under-densities at
$z\sim0.15$,	$0.3$ and $0.4$. In particular, the significant
under-density $0.35<z<0.5$ that occurs in both fields could point to a
selection effect in the survey. However, the mean GAMA redshift
distribution shown in Fig. \ref{fig:redshift_dist}(a) shows little
evidence for this. It also raises the question of whether or not some of
these features could be coherent between G23 and the Cold Spot.
Certainly at the lowest redshifts of $z < 0.05$ the under-density is
consistent with the `Local Hole' which spans the SGC
\citep{WhitbournandShanks}. In Section \ref{coherence} we shall use the
2dF Galaxy Redshift Survey (2dFGRS, \citealp{Colless}),  whose Southern
Strip spans the SGC between GAMA G23 and the Cold Spot, to check if this
apparent coherence is real or accidental. 

Meanwhile, the $n(z)$ similarities open up the possibility of G23 still acting as  a control field because it does not show a CMB Cold Spot. Therefore due  to the similarities in the 
redshift distributions of G23 and 2CSz we have fitted the density contrast 
in the same way as the
Cold Spot as shown in Fig. \ref{fig:void_fit}(b). The parameters of the
best fit are summarised in Table \ref{tab:void_params} with N=6 perturbations selected (see Table \ref{tab:AIC}), including 4 under-densities and 2 over-densities. We note that the highest redshift feature has been fitted with an under-density with an interior, narrower over-density which together fit the two $z>0.37$ under-densities seen in Fig. \ref{fig:void_fit}(b). The fitting procedure selects this over two under-densities because the under-densities are sharp and the density profile provides a poor fit individually. As we will discuss in Section \ref{systematics} we believe these features are affected by systematics and therefore we did not re-fit them.


\section{Discussion}

We have detected three large under-densities along the CMB Cold Spot
sightline, the largest with radius $r_0=119\pm35$h$^{-1}$Mpc centred at
$z_0=0.14$ with a central density contrast of $\delta_0=-0.34$. 
This supervoid is smaller but more under-dense than that proposed
by \cite{Szapudi} which has $r_0\sim220$h$^{-1}$Mpc and
$\delta_g=-0.25$. The \cite{Szapudi} void also has a higher central redshift
at  $z\sim0.22$ and may include the other 2CSz voids at $z_0=0.26$ and
$z_0=0.30$ (see Table \ref{tab:void_params}), seen as a single supervoid
due to the photo-$z$ errors. \cite{Kovacs} drew upon additional datasets
to suggest that the proposed supervoid extended back to zero redshift
with radius 500h$^{-1}$Mpc and with a smaller 195h$^{-1}$Mpc radius in 
the angular direction. From eq. \ref{eqn:dT} we estimate the
central temperature decrement due to our $z=0.14$ void at $-6.25\pm5.7\mu K$, small compared to some previous work (\citealp{Kovacs}), as expected due to the strong relationship between void radius and its ISW imprint. The combined ISW imprint of the three Cold Spot voids
is $-9.1\pm6.1\mu K$ and even adding the fourth questionable void this
rises to just $-31.7\pm15.9\mu K$. As we will discuss in section 
\ref{systematics} we believe the $z=0.42$ void is exaggerated by systematics.  
We also note that these estimates of the ISW 
imprint depend on the chosen void density profile used in the fitting process.
Although the profile used here (eq. \ref{eqn:void_density}) is not unique it is at least representative of what previous studies have done and allows for direct comparison with literature (e.g. \citealp{Kovacs}, \citealp{Finelli}).

The strongest evidence against an ISW explanation for the Cold Spot that may arise
from our results is due to the similarity in the $n(z)$ between GAMA G23
and the Cold Spot. Despite this, G23 has no CMB Cold Spot. Indeed, the predicted central ISW
decrement for G23 from summing the contributions in Table 
\ref{tab:void_params} above (excluding the features at $z>0.4$) is $-3.6\pm7.5\mu K$, 
statistically consistent with the $-9.1\pm6.1\mu K$
predicted similarly for the Cold Spot. The predicted central ISW
decrement for G23 is also consistent with that of the Cold Spot, even if no features in Table \ref{tab:void_params} are excluded. However, the CMB in the G23 sightline shows only a small central temperature decrement of $-15.4\pm0.3\mu$K, some
$\sim10\times$ lower than for the Cold Spot. Thus the similarity in the
large-scale structure between G23 and the Cold Spot fields forms a
further qualitative argument against foreground voids playing any
significant role in explaining the Cold Spot. On this evidence alone the
detected void cannot explain the CMB Cold Spot because a similar void in G23
has no such effect. 

\begin{figure}
	\centering
	\includegraphics[width=\linewidth]{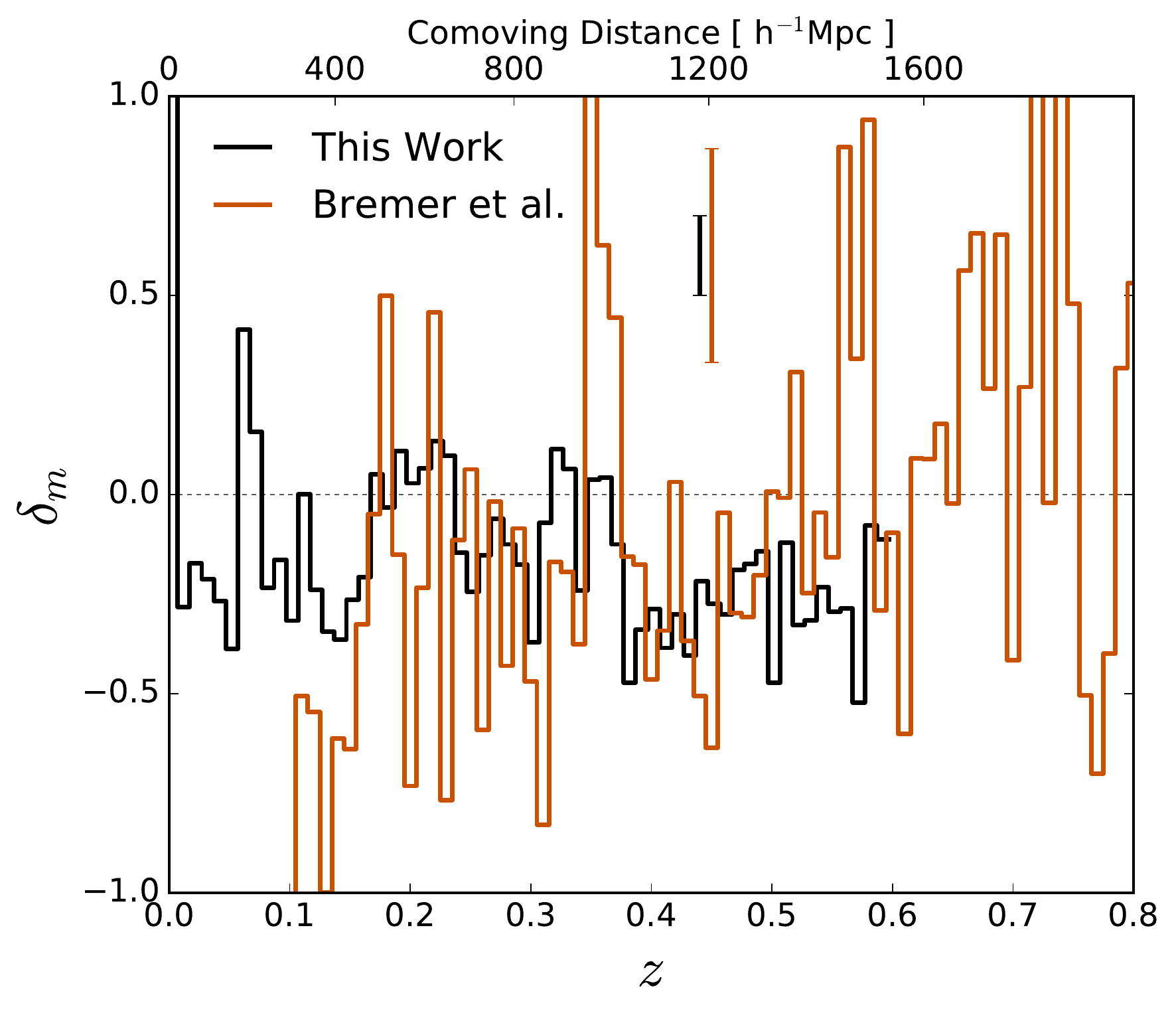}
	\caption{The 2CSz $\delta(z)$ (black) compared with VLT VIMOS
	\citep{Bremer} $\delta(z)$ (orange) to test the reproducibility of
	the Cold Spot void at $z=0.42$. Here a bias of $b=1.35$ has been
	assumed for the 2dF $\delta(z)$ and a bias of $b=1$ has been assumed
	for VLT VIMOS $\delta(z)$. Typical errors are plotted above the lines, Poisson errors are assumed for the VIMOS data.}
	\label{fig:Bremer}
	\vspace{-1em}
\end{figure}

\subsection{The reality of the $z=0.42$ void}
\label{systematics} 

In the Cold Spot $n(z)$ an apparent, relatively strong, void can be
seen at $0.37<z<0.5$ but we have already noted this is in a range where not only are
the statistics poorer but where we know that magnitude dependent 
incompleteness becomes more important. The similarity of this feature with the $0.34<z<0.5$ under-densities in G23 suggests there may be some sort of selection effect or systematic which we will now investigate. 

 We therefore test the
reality of this void in Fig. \ref{fig:Bremer} where we compare the 2CSz
$\delta_m(z)$ and the previous \cite{Bremer} VLT VIMOS $\delta_m(z)$ and see
that an under-density at $z=0.42$ may also be detected in that dataset,
albeit at low $\sim2\sigma$ significance. A lower bias of $b=1$ has also
been assumed here for the VIMOS $\delta_m(z)$ compared to $b=1.35$ for 2CSz,
on the grounds that the VIMOS galaxies are intrinsically fainter. 
This is consistent with results from the VVDS survey \citep{VVDScluster}. 
We note that despite this apparent agreement the VLT VIMOS data probes 
a much smaller volume at this low redshift end and therefore would have 
large sample variance.

The absence of this feature from the mean GAMA $n(z)$ indicates that this feature cannot be intrinsic to the $i_{AB}<19.2$ selection criteria. We instead
suggest that it may be due to a systematic selection effect.
 Although
the other GAMA fields are apparently unaffected by this systematic,
this may be explained by the Cold Spot (and G23) data having slightly
lower S/N due to somewhat shorter 2dF exposure times and redshift
success rate viz. Cold Spot (45mins, 89.0$\%$), G23 (30-50mins,
94.1$\%$) vs. the other 3 Equatorial GAMA fields used here (50-60 mins, 98.5$\%$).
2CSz was also conducted in gray time which will further reduce the S/N with respect to GAMA. The lower S/N ratio will increase spectroscopic incompleteness
and we note that the 4000\r{A} break and Ca II H and K absorption lines transition though
the dichroic over this redshift range while the H$\alpha$ emission line also leaves
the red arm of the spectrograph. It is possible that these two effects
make accurate redshifting more difficult over this redshift range and
would create an apparent under-density.  To test this we split 2CSz into
pointings with high and low spectroscopic success rate, with half having
a success rate greater than $90\%$ and half with less. The result of
this is shown for $z\ge0.38$ in Fig. \ref{fig:void_fit}(a) by the dashed histogram. 
All fitting used the full dataset. The success
rate of the 2dF field strongly affects the depth of the $z=0.42$ void
indicating that it is affected by systematic incompleteness. 

Also, at $z>0.4$,  small differences in the homogeneous model will lead
to large differences in the derived $\delta_m(z)$. To investigate whether
the model $n(z)$ could be over-predicting the galaxy density at the
higher redshifts creating spurious under-densities, we have explored
a model $n(z)$ constructed from random catalogues built for the GAMA 
survey \citep{Farrow} and find that indeed this
different model $n(z)$ decreases the depth of the $z=0.42$ void. 
When compared to the mean GAMA $n(z)$ however this model 
$n(z)$ appears to under-predict the galaxy density at higher redshift and 
therefore we do not replace our homogenous model with the GAMA random 
catalogue constructed $n(z)$. Whether
the void seen by 2CSz in this $z$ range is accentuated by such
systematics or not does not matter for our main conclusion since even
including this void's contribution the total ISW decrement from Table
\ref{tab:void_params} is still only $\sim-32 \mu$K  compared to the 
$\sim-150 \mu$K needed to explain the Cold Spot.

Additionally we note that the bias of galaxies will not be constant throughout the redshift range as assumed. Because the survey is magnitude limited the galaxies at the high redshift end of the survey will be brighter than the low redshift end. The brighter 2CSz galaxies at $z=0.42$
may actually be as large as $b\sim2$ (e.g. \citealp{Zehavi2011}) and increasing the bias would
linearly decrease the depth of the void $\delta_0$ (by eqn \ref{eqn:matter_density_contrast}) and hence its ISW imprint. 

Together these arguments cast doubt on the existence of the $z=0.42$ void and for this reason we neglect it in our conclusions. A sample of galaxies with a magnitude limit intermediate between that of 2CSz and \cite{Bremer} et al is needed  to determine finally the status  of the $z=0.42$ void.

\begin{figure}
	\centering
	\includegraphics[width=\linewidth]{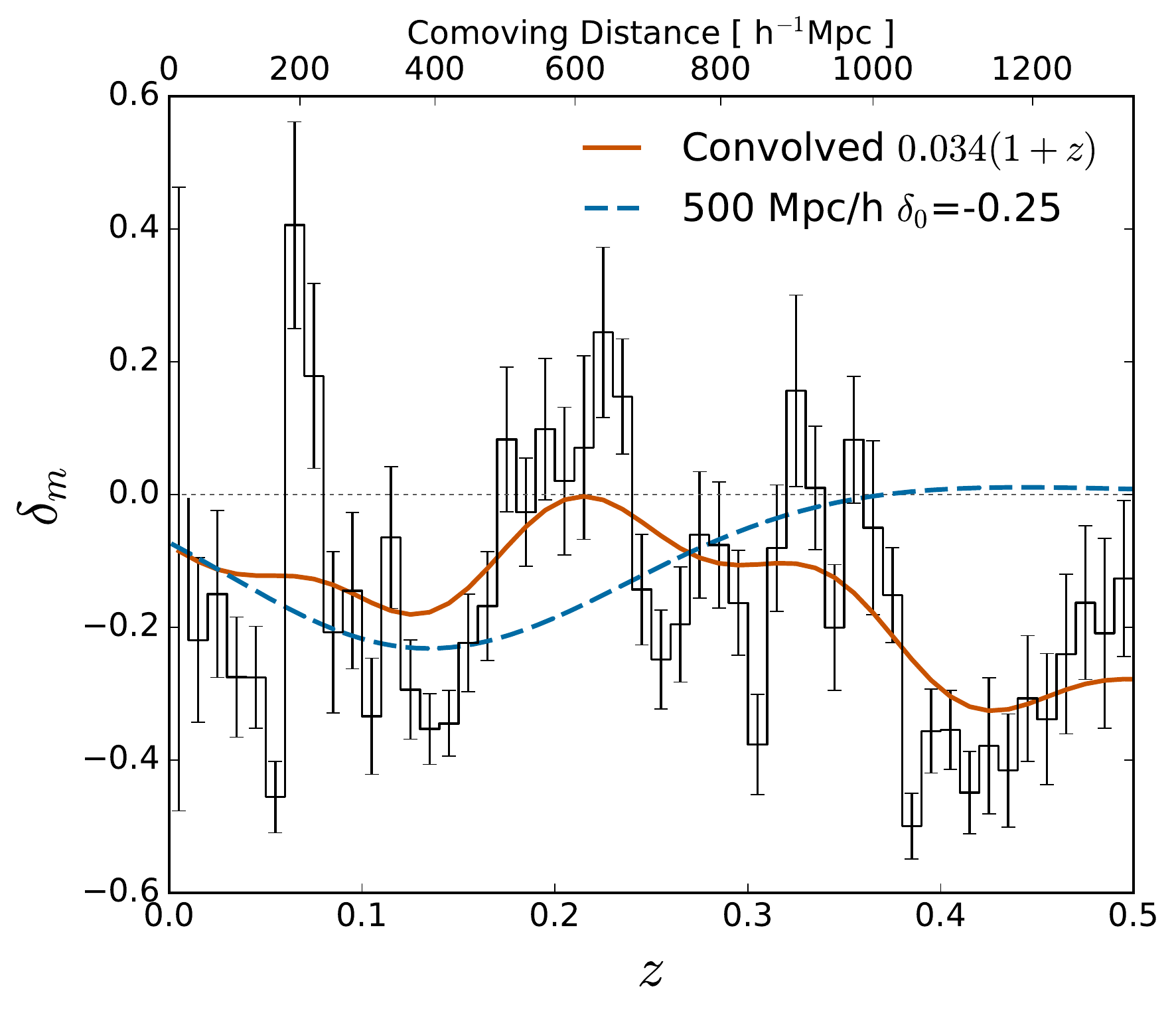}
	\caption{The 2CSz $\delta_m(z)$ (black), the 2CSz $\delta_m(z)$ convolved with the photo-$z$ error of the PanSTARRS data of \citet{Szapudi} (orange) and compared to the fitted $\delta_m(z)$ model of \citet{Kovacs} (blue) }
	\label{fig:convolve}
	\vspace{-1em}
\end{figure}

\subsection{Photo{\it -z} and spectroscopic $n(z)$ }
\label{photoz} 

In order to assess why the spectroscopic 2CSz survey results apparently differ
from the photometric redshift survey of \cite{Kovacs} we convolved the
2CSz spectroscopic redshift distribution with an estimated error of
$0.034(1+z)$ photo{\it -z} error, which is the quoted photo-$z$ error from \cite{Szapudi}. \cite{Kovacs} used 2MPZ with a very small photo-$z$ error of $0.015(1+z)$,
but the 2MPZ sample is limited by low number densities at higher redshifts so we do not compare to this directly.

The resulting model $\delta_m(z)$ is
shown in Fig. \ref{fig:convolve} where we see that there is limited
consistency with the model result of \cite{Kovacs} with $r_0=500$h$^{-1}$Mpc
and $\delta_0=-0.25$ when convolved with a photo-$z$ error. The main source of disagreement is the lack of an 
under-density at $z\sim0.2$ in 2CSz which seems difficult to reconcile with the model void but 
we note that at $z>0.15$ the 2MPZ data is consistent with no under-density due to a large uncertainty. While our data is not consistent with an $r_0=500$h$^{-1}$Mpc void we believe it is consistent with the photo-$z$ data.

When we compare our predicted ISW central decrement to previous work we see some consistency. With the 3-void model of the Cold Spot line of sight the combined temperature decrement is $-9.1\pm6.1\mu K$ which is consistent with the $\sim -20 \mu K$ of \cite{Szapudi} but not with the $\sim -40 \mu K$ of \cite{Kovacs}. One could argue the 4-void model at $-31.7\pm16.0 \mu K$ is consistent with \cite{Kovacs} values, but $\sim \frac{2}{3}$ of that decrement is due to the $z=0.42$ void which is likely to be contaminated by systematic effects as discussed previously. Additionally the void of \cite{Kovacs} did not extend to $z>0.4$ and it is beyond the range of the 2MPZ data.


\begin{figure}
	\centering 
	\includegraphics[width=\linewidth]{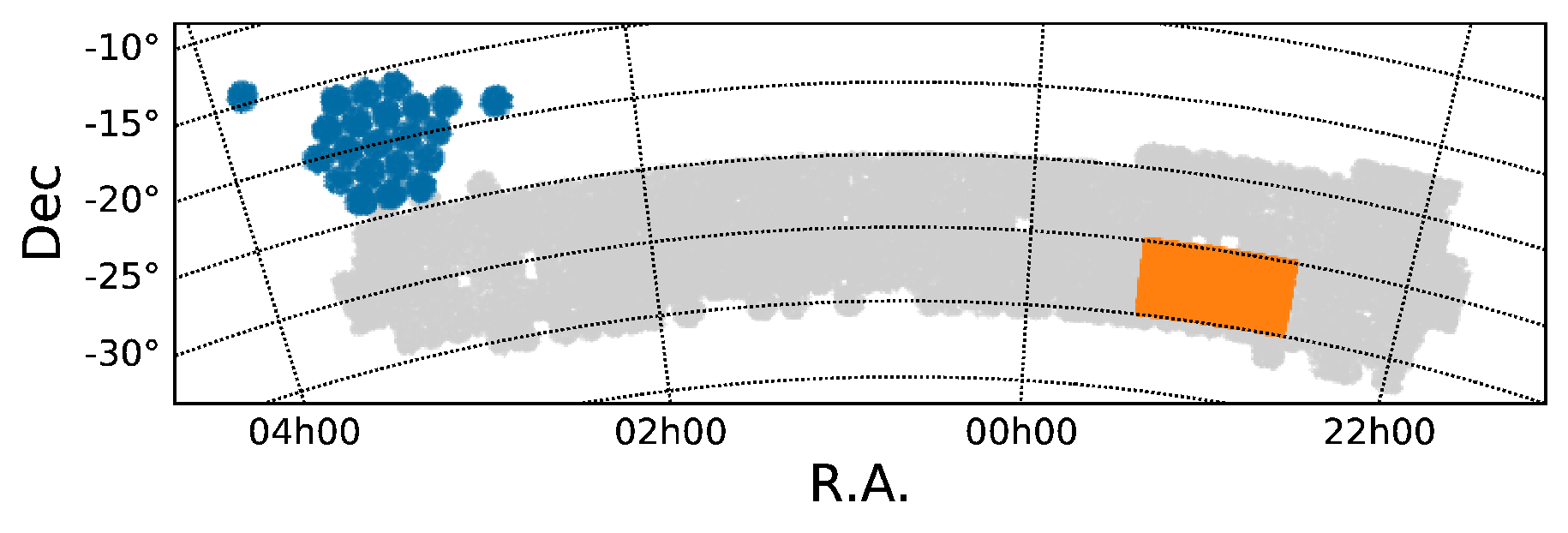} 
	\caption{The  position of the 2dFGRS SGC strip 	(grey) relative to 
	the 2CSz 2dF fields (blue) and the GAMA G23 area (orange). }
	\label{fig:2dFGRS_CS_G23}
\end{figure}

\subsection{A coherent SGC galaxy distribution?}
\label{coherence} 

We have already discussed the important question of the normalisation
of the Cold Spot $n(z)$. Both G23 and the Cold Spot areas are contained in the Local
Hole under-density known to extend at least to $z=0.06$ across the SGC.
Moreover we have noted that the galaxy count in the 5$^{\circ}$ radius Cold Spot
area is $\sim16$\% under-dense relative to G23 and the rest of the SGC at
our $i_{AB}<19.2$ limit. When compared to a surrounding  $\sim1000$deg$^2$ area the 5$^{\circ}$ core of the Cold Spot is $7.4\%$ under-dense. The Cold Spot area therefore appears to exist
in an  environment exhibiting a significant global  gradient stretching
across the SGC. Finally we have noted the similarity of the 2CSz and GAMA
G23 redshift distributions which again may suggest evidence for coherent
structure extending between them.

\begin{figure*}
	\centering
	\includegraphics[width=\textwidth]{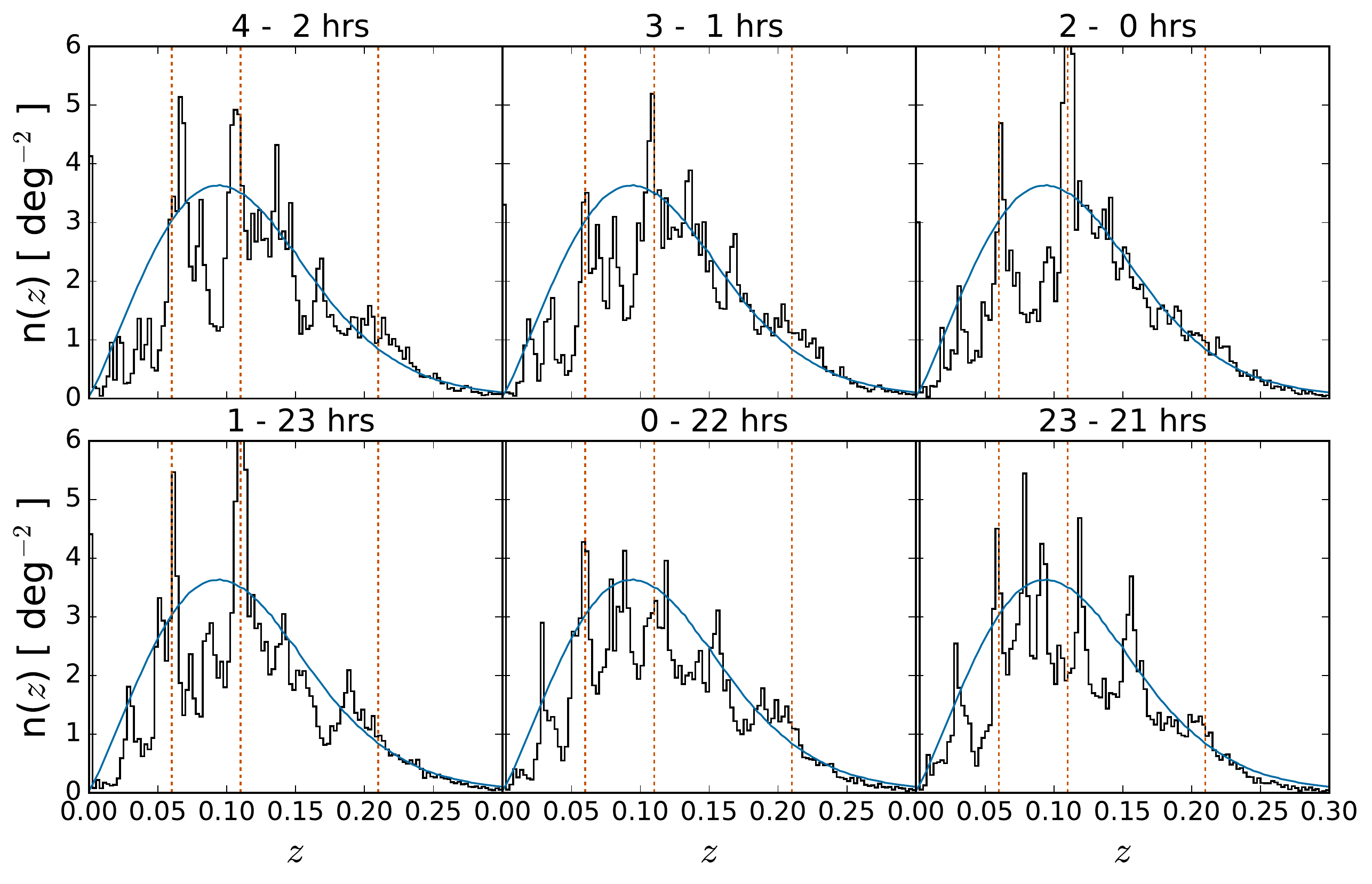}
	\caption{The 2dFGRS SGC galaxy redshift distributions, $n(z)$  in
	overlapping 2hr ranges of R.A. at Dec$\sim-30\pm5$deg (black). The homogeneous model prediction of \citet{Metcalfe} to the 2dF
	limit of $b_j=19.6$ is plotted (blue). The redshifts corresponding to the peaks in
	the average 2dFGRS $n(z)$ at $z=0.06$, 0.11 and 0.21 are marked (orange dashed lines). 
	 }
	\label{fig:2dFGRSnz}	
\end{figure*}

To investigate further this possibility, we now exploit the 2dF Galaxy
Redshift Survey (2dFGRS, \citealp{Colless}) which spans the SGC
between GAMA G23 and the Cold Spot at $-35^\circ<$ Dec  $<-25^\circ$
(see Fig. \ref{fig:2dFGRS_CS_G23}). With a magnitude limit $b_J (\sim g)
\le19.6$, 2dFGRS is shallower than the $i_{AB}\le19.2$ surveys so only probes
the low $z$ structures but has a large area. \cite{Busswell} shows
the redshift distribution of the 2dFGRS survey in the SGC in their Fig.
14 (also shown in \citealp{Norberg2002} Fig. 13). The distribution shows peaks at $z=0.06$, $z=0.11$ and $z=0.21$ which
are very similar to those shown in 2CSz and roughly similar to those
shown in G23. We have attempted to track these features across 2dFGRS to 
see if they do in fact span the sky between G23 and 2CSz. 
When we split 2dFGRS by R.A. as in Fig.
\ref{fig:2dFGRSnz}	we generally see coherence in that at $z<0.06$ we
consistently see under-density in this range. This is the `Local Hole' of
\citet{WhitbournandShanks} (see their Fig. 2b) which covers $\sim3500$ deg$^2$
of the SGC (the 6dFGS-SGC area marked in orange in their Fig. 1 with
coordinate ranges given in their Table 3). Based on the $0.06<z<0.11$
void seen in the 2dFGRS $n(z)$ shown in Fig. 14 of \citet{Busswell},
these authors have speculated that the void runs to $z\sim0.1$.  In passing, we note that the $\sim8\%$ gradient between the regions surrounding G23 and the Cold Spot may represent Local Hole sub-structure. 

In Fig. \ref{fig:2dFGRSnz} we see that the eastern half of 2dFGRS ($0<$
R.A. $<4$hrs) more clearly exhibits the peaks at $z=0.06$ and $z=0.11$
(with intervening under-density) than does the range at $21<$ R.A. $<0$hr.
We have checked that restricting 2dFGRS to the G23 area produces very good
agreement in $\delta_m(z)$ out to $z<0.25$. More speculatively, even the
 $z=0.21$ peak may be seen in at least some of the R.A. ranges If so, 
this possible coherence may also explain why 2CSz and G23 have such
similar $n(z)$ distributions. However in the $23<$ R.A. $<1$hr and $0<$ R.A. $<2$hr ranges the feature at $z=$0.21 is less obvious and  perhaps argues against coherence extending to $z\sim0.2$. This would leave the similarity of the 2CSz and G23  $n(z)$'s at $0.1<z<0.2$ appearing accidental.  We note that the absence of these structures from the NGC 2dFGRS survey (c.f. Figs. 13,  14 of \citealp{Busswell}) makes systematic effects unlikely as the cause.

How likely is it, in the standard cosmological model, that coherent 
structure extends out to $z < 0.2$ across the 2dFGRS SGC strip? We assume 
an $\sim$1000deg$^2$ area for 2dFGRS SGC and a power-law correlation 
function, $\xi(s)=(s/s_0)^{-\gamma}$, with $s_0\sim6.92$h$^{-1}$Mpc and 
$\gamma\sim1.51$ for $s<50$h$^{-1}$Mpc, as measured for 2dFGRS by 
\citet{Hawkins2003}. The variance, $\sigma_N^2$, of galaxy numbers, $N$, 
around average $\bar{N}$ in a volume, $V$, where the galaxy space density, 
$n$ $(=N/V)$, is (e.g. \citealp{Peebles1980})

\begin{equation}
\centering 
\sigma_N^2  =\langle(N-\bar{N})^2\rangle=\bar{N}+n^2\int_V\xi(s_{12})dV_1dV_2 ,
\label{eqn:sigma}
\end{equation}

\noindent implying $\sigma_N\sim20\times\sqrt{\bar{N}}$. Given that 
$\bar{N}\sim140000$ galaxies in the 2dFGRS SGC volume, a nominal 10\% 
under-density (or over-density) across 2dFGRS SGC even out to $z\sim0.2$
would amount to a $\sim1.9\sigma$ fluctuation. On the same assumptions, a
similar over- or under-density out to $z=0.1$ would represent a
significance of $\sim1.3\sigma$.  Now these may  be taken as a rough
measure of the significance of coherence in a survey modeled by some of
its $z$ range being 10\% overdense and the rest being 10\% underdense.
So at $\sim1.3-1.9\sigma$, we conclude that galaxy clustering
coherence  across 2dFGRS SGC can plausibly explain the 2CSz-G23
coherence out to $z\sim0.1$ and more speculatively to $z\sim0.2$. 
However the observational evidence for coherence at $z\sim0.2$ is mixed.

\subsection{Origin of The CMB Cold Spot}

\begin{figure}
	\centering
	\includegraphics[width=\linewidth]{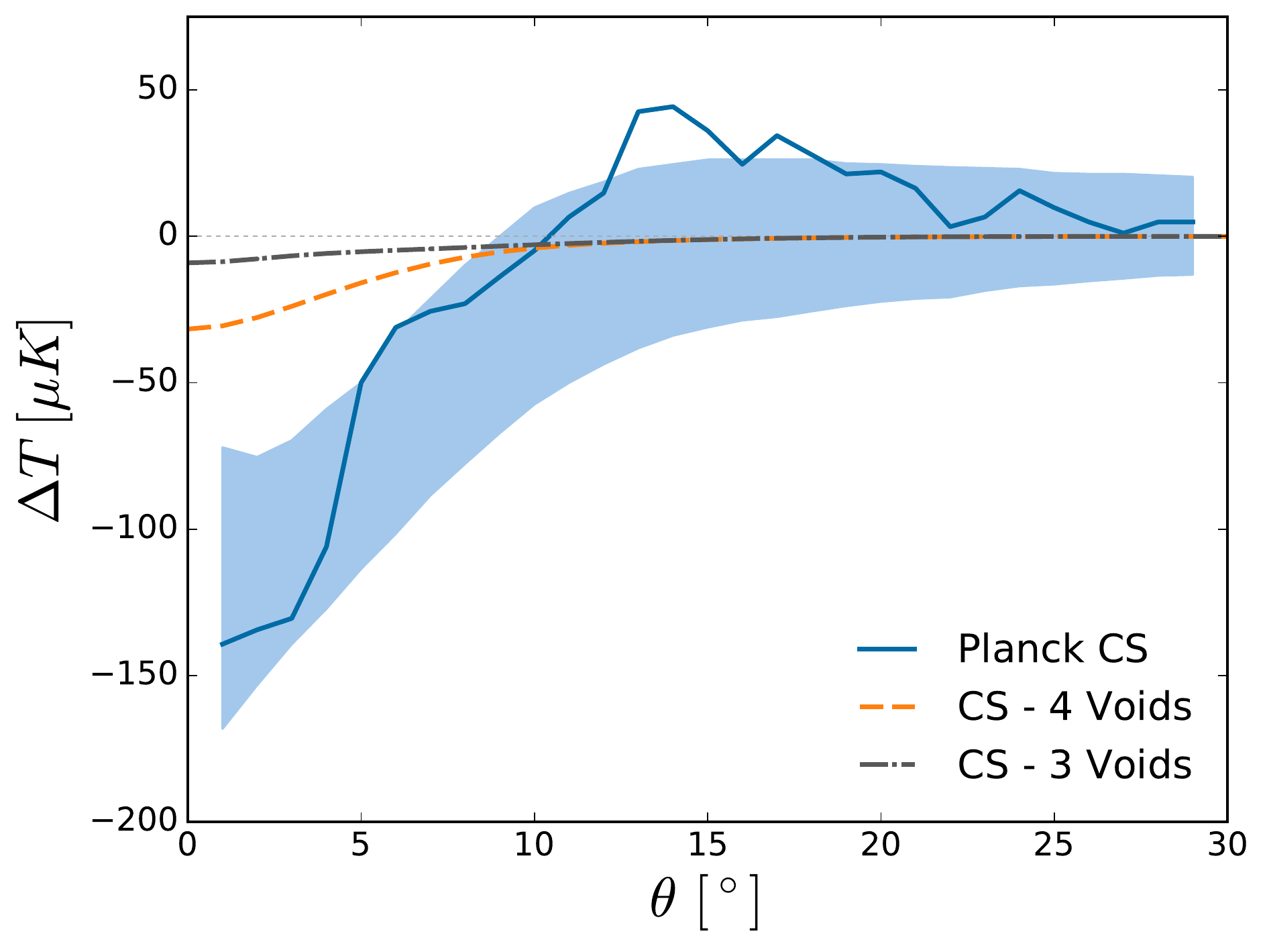}
	\caption{The Cold Spot temperature profile \citep{Planck2016CS} (blue line) and the ISW imprints of the 3- and 4-void models (grey dot-dashed and yellow dashed respectively) fitted to the Cold Spot region.
	The void temperature profiles from Table \ref{tab:void_params} have been summed and the
	result fitted to eq. (3) of \citet{Naidoo}. The shaded region (light blue) is the $68\%$ confidence interval from the coldest spots identified in Gaussian simulations (see \citealp{Nadathur}, Fig 6).}
	\label{fig:CS_profile}
	\vspace{-1em}
\end{figure}

As noted in Section \ref{sec:intro}, several authors have calculated the
significance of the Cold Spot with respect to the coldest spots in CMB
sky simulations (e.g. \citealp{Nadathur}, \citealp{Planck2016CS}). The
significances are typically at the $\sim1$\% level. As shown by these
authors, the significance of the Cold Spot in the standard cosmology
comes not from the central temperature but from the temperature profile
seen in Fig. \ref{fig:CS_profile} which closely matches the compensated
SMHW that was originally used to detect it (\citealp{Vielva}).  On this
basis when assessing what impact the detected voids have on the
significance of the Cold Spot we have to go beyond central temperature
and look at  the significance of the SMHW filtered temperature
subtracted for the detected voids. This removes the ISW imprinted signal
and assesses the significance of the residual primordial profile.
Following \cite{Naidoo}, subtracting our best 3-void (i.e. the voids
with $z_0<0.4$ in Table \ref{tab:void_params}) model ISW contribution
would reduce the significance of the Cold Spot only slightly, typically
to $\sim1.9\%$ (\citealp{Naidoo}) i.e. only 1 in $\sim50$ $\Lambda$CDM
Universes would produce such a feature by chance. Fig.
\ref{fig:CS_profile} shows the ISW imprints of
the 3 and 4-void models and the measured CMB Cold Spot temperature
profile. This significance would be reduced if our 4-void
model was trusted but, as previously argued, the void at $z=0.42$ may be
unduly affected by systematics. 

\citet{Kovacs} claimed the Cold Spot supervoid is an elongated supervoid  at $z=0.14$ with
$r_0=500$h$^{-1}$Mpc in the redshift direction and $r_0=195$h$^{-1}$Mpc
in the angular direction with $\delta_0=-0.25$. The
ISW effect on the central decrement is estimated to be a reduction of $\sim40\mu$K.
At the central redshift of $z=0.14$ this supervoid would extend $27.5^{\circ}$ on the sky.   
We note
that the 2dFGRS SGC strip covers the area to the South of the Cold Spot.
In the 2h$<$R.A.$<$4h range, all of this R.A. bin is within $27.5^{\circ}$ of the Cold Spot. Fig. \ref{fig:2dFGRSnz} shows that although there is a 2dFGRS void at $z=0.08$ within the 
supervoid  redshift range, the peak at $z=0.11$ and plateau out to $z=0.15$
 is near the claimed $z=0.14$
centre  of the supervoid; there seems little evidence of a void
at $0.1<z<0.25$ in this 2dFGRS 2h$<$R.A.$<$4h range. The $z=0.2$
peak may still be present indicating there may be an under-density at $0.15<z<0.2$. 
So at least in the direction South of the
Cold Spot, evidence for an extended simple void structure around its
centre is again not present.


Various authors (e.g. \citealp{Cai2014fr,Cai2014imprint,KovacsDES} and
references therein) have also discussed the possibility of an enhanced
ISW effect in voids being produced by modified gravity models. This has
been done to explain observations where a  larger than expected 
($2-4 \times$ under $\Lambda$CDM) ISW-like signal has been found around voids (\citealp{Granett}, \citealp{Cai2016}), these results are however low significance. 
It may be speculated whether our 2CSz Cold Spot results may also
be explained similarly. But again the similarity between the galaxy redshift
distributions in 2CSz and the G23 control field tends to argue against
this possibility. If some modified gravity model did give an enhanced ISW
effect to explain the Cold Spot then why is there no similar Cold Spot
seen in the G23 line-of-sight? This argument should be
tempered with the facts  that, first, the $n(z)$ agreement between the
Cold Spot and G23 is inexact given that the $n(z)$ peak at $z=0.21$ is more
pronounced in G23. This difference is reflected in 
the predicted ISW decrements, $-9.1\pm6.1\mu K$ and $-3.6\pm7.5\mu K$ for the Cold Spot and G23 respectively. 
Second, the $n(z)$'s used to construct the $\delta_m(z)$'s were
normalised with respect to their surroundings and so don't contain all
the information of the largest scale fluctuations. As discussed 
previously the region surrounding the Cold Spot is under-dense with 
respect to the region surrounding G23 by $\sim8\%$ so the two fields are not exactly 
equivalent and the structures detected in this analysis are 
embedded in different large scale potentials. 
This could have an effect on the Cold Spot ISW imprint but likely 
at larger scales than the 5$^{\circ}$ radius feature we
have mainly investigated here. One could argue that the alignment of 
the CMB Cold Spot and the large $z=0.14$ void implies a causal link 
though the improbability of alignment but voids of this 
scale are not expected to be unique (\citealp{Nadathur}, 
\citealp{KovacsDES}) and our search was not blind nor the 
only attempt to detect for a void.

If not explained by a $\Lambda$CDM ISW effect the Cold Spot could 
have more exotic primordial origins. If it is a non-Gaussian feature, 
then explanations would then include either the presence
in the early universe of topological defects such as textures
(\citealp{CruzTexture}) or inhomogeneous re-heating associated with
non-standard inflation (\citealp{BuenoSanchez}). Another explanation
could be that the Cold Spot is the remnant of a collision between our
Universe and another `bubble' universe during an early inflationary
phase (\citealp{ChangKlebanLevi}, \citealp{LarjoLevi}). 
It must be borne in mind that even without a supervoid the Cold Spot
may still be caused by an unlikely statistical fluctuation in the
standard (Gaussian) $\Lambda$CDM cosmology.


To conclude, based on the arguments and caveats above we have ruled out 
the existence of a void at which could imprint the majority of the CMB Cold Spot 
via a $\Lambda$CDM ISW effect. The predicted decrement is consistent with some previous studies \citep{Szapudi}, although certainly at the low end of literature values. 
We have additionally placed powerful constraints on any non-standard ISW-like effect which must now show how voids, apparently unremarkable on $5^{\circ}$ scales, can imprint the unique CMB Cold Spot. 
The presence of the detected voids only slightly relaxes the significance of the primordial residual of the CMB Cold Spot in standard cosmology to approximately 1 in 50, tilting the balance towards a primordial and also possibly non-Gaussian origin. But at this level of significance  clearly  any exotic explanation will have to look for further evidence beyond the Cold Spot temperature profile.

\section{Conclusions}

We have conducted a spectroscopic redshift survey of the CMB Cold Spot core in order to test claims from photo-$z$ analyses for the existence of  a large low-$z$ void that could be of sufficient scale and rarity to explain the CMB Cold Spot.

$\bullet$ We have detected an $119$ h$^{-1}$Mpc, $\delta_{g}=-0.34$
under-density at $z=0.14$. This under-density is much less extended than
found in photo{\it-z} analyses in the literature but is more under-dense. The
estimated $\Lambda$CDM ISW effect from this void is estimated at $-6.25
\mu$K, much too small to explain the CMB Cold Spot.

$\bullet$ Two further small under-densities were observed at $z=0.26$
and $0.30$. The effect of these voids is even smaller than the
$z=0.14$ void.

$\bullet$ A further candidate void was detected at $z=0.42$ although we
conclude this is most likely due to redshift incompleteness in the
survey. Even if real this void would still not explain the CMB Cold Spot.

$\bullet$ Without detailed calculation we have shown that the rarity of
this void is not sufficient to motivate it as the cause of the CMB
Cold Spot because of the similarity with GAMA G23. The comparability of
under-densities at  $z\sim0.4$ between G23 and the Cold Spot again means that
even if the $z=0.42$ void in the Cold Spot was not a systematic effect,
it is not unique enough to suggest an effect beyond standard cosmology.

$\bullet$ Combining our data with previous work \citep{Bremer} 
the presence of a very large void which can explain the CMB Cold Spot
can be excluded up to $z\sim1$, beyond which the ISW effect becomes
significantly reduced as the effect of the Cosmological Constant is
diluted. 

$\bullet$ The similarity between the  2CSz and G23 $n(z)$ distributions
may have some explanation in the similar $n(z)$ seen in the 2dFGRS SGC
strip that spans the $\sim60^\circ$ angle between these sightlines. This includes
the `Local Hole' at $z<0.06$ but may also include further  structures out to
$z\sim0.2$. 

Our 2CSz results therefore argue against a supervoid explaining 
a significant fraction of the Cold Spot via the ISW effect. This suggests 
a primordial origin for the Cold Spot, either from an unlikely 
fluctuation in the standard cosmology or as a feature produced by  
non-Gaussian conditions in the early Universe.

\section*{Acknowledgements}
We thank N. Metcalfe, K. Naidoo and A.J. Smith for valuable discussions.
We acknowledge the GAMA team for providing survey data in advance of publication. 
Based on data products from observations made with ESO Telescopes at the
La Silla Paranal Observatory under program ID 177.A-3011(A,B,C,D,E.F).
We further thank OPTICON and the staff at the Australian Astronomical
Observatory for their observing support at the AAT. GAMA is a 
joint European-Australasian project based around a spectroscopic
campaign using the Anglo-Australian Telescope. The GAMA input catalogue is
based on data taken from the Sloan Digital Sky Survey and the UKIRT Infrared
Deep Sky Survey. Complementary imaging of the GAMA regions is being obtained
by a number of independent survey programmes including GALEX MIS, VST KiDS,
VISTA VIKING, WISE, Herschel-ATLAS, GMRT and ASKAP providing UV to radio
coverage. GAMA is funded by the STFC (UK), the ARC (Australia), the AAO, and
the participating institutions. The GAMA website is http://www.gama-survey.org/ .
The G23 GAMA dataset is in part based on observations made with ESO Telescopes
at the La Silla Paranal Observatory under programme ID 177.A-3016. PN acknowledges the
support of the Royal Society, through the award of a University Research
Fellowship and the European Research Council, through receipt of a
Starting Grant (DEGAS-259586). RM, TS \& PN acknowledge the support of
the Science and Technology Facilities Council (ST/L00075X/1 and ST/L000541/1). MLPG
acknowledges CONICYT-Chile grant FONDECYT 3160492.



\bibliographystyle{mnras}
\bibliography{bibfile}




\bsp	
\label{lastpage}
\end{document}